\documentclass[prl,twocolumn,superscriptaddress]{revtex4}
\usepackage{epsfig}
\usepackage{times}
\usepackage{amsmath}
\usepackage{amsfonts}
\usepackage{amssymb}
\usepackage[usenames]{color}
\usepackage{graphicx}
\newcommand{\SCB}[1]{\textcolor{black}{#1}}
\newcommand{\NTT}{NTT Basic Research Laboratories, NTT Corporation, 3-1 Morinosato-Wakamiya, Atsugi, Kanagawa, 243-0198, Japan.}

\newcommand{\oxford}{Department of Materials, University of Oxford, OX1 3PH, United Kingdom}

\newcommand{\beq}{\begin{equation}}
\newcommand{\eeq}{\end{equation}}
\newcommand{\beqa}{\begin{eqnarray}}
\newcommand{\eeqa}{\end{eqnarray}}

\begin{document}
\title{
Magnetic field sensing
with quantum
error detection
under the effect of energy relaxation}
\author{Yuichiro Matsuzaki}
   \affiliation{
\NTT
   }
 \author{Simon Benjamin
 }
 \affiliation{\oxford}

\begin{abstract}
 A solid state spin is an attractive system \SCB{with which} to realize an
 ultra-sensitive magnetic field sensor.
 \SCB{A spin superposition state will acquire a phase induced by the target field, and we can estimate the field strength from this phase. Recent studies have aimed at improving sensitivity through the use of quantum error correction (QEC) to detect and correct any bit-flip errors that may occur during the sensing period.} 
Here, we investigate the performance of a two-qubit sensor \SCB{employing QEC and} under the effect of energy
 relaxation. Surprisingly, we \SCB{find} that the standard QEC technique to
 detect and recover \SCB{from an} error does not improve the sensitivity compared
 with the
 \textcolor{black}{single-qubit sensors}.
 This is a consequence \SCB{of} the fact that
 the energy relaxation induces both a phase-flip and a bit-flip noise
 where the former noise cannot be distinguished from the relative phase
 induced from the target fields. However, we have found that
 we can improve the sensitivity if we adopt
 postselection to discard the state when error is detected.
 \SCB{Even when quantum error detection is moderately noisy, and allowing for the cost of the
 postselection technique, we find that this two-qubit system shows an advantage in 
 sensing over a single qubit in the same conditions.}
\end{abstract}

\maketitle


Measurement of small magnetic fields is an important objective in the field of
metrology because of many practical applications in material science, biology, and
medical science. It is known that SQUID \cite{simon1999local}, Hall
sensors \cite{chang1992scanning}, and force sensors \cite{poggio2010force}
show excellent performance for such field detection.

A two-level system coupled
with magnetic fields is an alternative way to detect small magnetic
field.
The magnetic field typically shifts the resonant frequency of the qubit,
and one can readout the shift from Ramsey type interference experiment,
to which we refer as \textcolor{black}{single-qubit sensors}.
Atomic vapor magnetometry is one of such ways to use a qubit for
sensing \cite{budker2007optical}.
Nitrogen vacancy centers in diamond is another candidate to realize such
a sensor \cite{maze2008nanoscale, taylor2008high, balasubramanian2008nanoscale},
and this  typically has a better spatial resolution compared with
other conventional devices.

One of the obstacles to sense small
magnetic fields with the qubit is decoherence on the system \cite{huelga1997improvement}.
The frequency shift from the magnetic fields induces a relative phase between
coherent superpositions of the qubit, and this provides us with
measurable signals \cite{degen2016quantum}. This means that any
deteriorating effect of the phase coherence 
decreases the signal to noise ratio of the sensing.
Since unwanted coupling with environment is inevitable, the decoherence
effect is one of the main challenges to realize an ultra-sensitive field sensor with qubits.

Recently, magnetic field sensing with quantum error correction (QEC) has been
proposed to improve the sensitivity of qubit-based metrology \cite{kessler2014quantum,arrad2014increasing,dur2014improved,herrera2015quantum}.
QEC is a technique to detect and recover errors by
using an encoded state where ancillary qubits are used to employ
redundancy in the code space \cite{gottesman2009introduction}. QEC has been proposed in the context of
scalable quantum computation,
and proof of principle experiments have been demonstrated in many systems
such as superconducting qubits
\cite{gladchenko2009superconducting,reed2012realization,corcoles2015demonstration,kelly2015state},
NV centers \cite{waldherr2014quantum}, and ion traps \cite{nigg2014quantum}.
Moreover, 
previous researches show that
QEC can be applied to enhance the sensitivity in the quantum
metrology with certain conditions \cite{kessler2014quantum,arrad2014increasing}.
Interestingly, even for a two-qubit system, it is in principle possible to enhance
the sensitivity by QEC if one of the qubits has much longer coherence
time than the other \cite{kessler2014quantum,arrad2014increasing}.
It is worth mentioning that we should not apply QEC to protect the qubit from the
dephasing during the sensing, because the relative
phase from the target magnetic field is indistinguishable from the unwanted phase induced by
the environment.
On the other hand, if a bit-flip
noise is relevant on the qubits,
quantum field
sensors with QEC can beat the
\textcolor{black}{single-qubit sensors}
\cite{kessler2014quantum,arrad2014increasing,dur2014improved}.
\SCB{Indeed, an experiment has been reported} where
such an enhancement of the sensitivity by QEC was demonstrated
under the effect
of artificial bit-flip noise by using a nitrogen vacancy center in
diamond or an optical setup
\cite{unden2016quantum,cohen2016demonstration}. However, \SCB{to our knowledge there has not yet been any} experimental
demonstration \SCB{of} the enhancement of the sensitivity of quantum
metrology by QEC \SCB{versus}
natural decoherence from the environment.

In this paper, we investigate the performance of the quantum field
sensing with QEC technique under the effect of energy relaxation.
A solid state spin qubit is affected by two type of decoherence, dephasing and
energy relaxation. The dephasing time of the qubit is characterized by
$T_2$ while the energy relaxation time is characterized by $T_1$ \cite{slichter2013principles}.
It is worth mentioning that dynamical decoupling \SCB{techniques are} available
to suppress the effect of the dephasing, \textcolor{black}{which can
improve the sensitivity for AC magnetic fields}
\cite{viola1999dynamical,taylor2008high}. 
With \SCB{a} well-controlled dynamical decoupling technique, the coherence time of the solid
state qubit can be in principle limited by the energy relaxation process, which
is observed in several systems such as superconducting qubits \cite{bylander2011noise,yan2015flux}.
However, the energy relaxation induces not only bit-flip noise but also
phase-flip noise. As pointed out in
\cite{kessler2014quantum,arrad2014increasing,dur2014improved},
QEC \SCB{versus} dephasing cannot be applied to
enhance the sensitivity of quantum metrology, because QEC erases
not only the environmental unknown phase but also the relative phase
induced by the target fields. So it is not \SCB{trivially obvious} if QEC can
improve the sensitivity if the energy relaxation is a relevant source of
the decoherence. \SCB{Here we} investigate the performance of quantum field
sensors using QEC, where the error is detected and recovered
on the encoded state with an ancillary qubit. \SCB{We} show that the
standard QEC approach does not improve the sensitivity over the
\textcolor{black}{single-qubit sensors}.
\SCB{We proceed to} consider a postselection strategy where
 we discard the state when the bit-flip error is
detected.
\textcolor{black}{Since we need to wait until we have successful
operations, the postselection effectively decrease the time resource.}
Interestingly, \textcolor{black}{even if we take into account the time
loss due to the postselection}, we show that such a postselection strategy
actually improves the sensitivity, and beats \textcolor{black}{single-qubit sensors}.
\SCB{Moreover, this is true even when the detection process is imperfect and can itself introduce noise.}
Our results show that an encoded two-qubit state is actually beneficial
\SCB{for} ultra-sensitive magnetic field \SCB{sensing} in realistic conditions.

\section{Magnetic field sensing with the standard Ramsey type sequence}
Let us review the standard Ramsey type sequence to estimate the magnetic fields with a
qubit \cite{huelga1997improvement}, \SCB{which we refer to as} \textcolor{black}{single-qubit sensors}.
Suppose that we have a qubit coupled with magnetic field, and the
Hamiltonian is described by
\begin{eqnarray}
 H=\frac{\omega }{2}\hat{\sigma }_z
\end{eqnarray}
where $\omega $ denotes a detuning due to the magnetic field $B$, and we
assume that the detuning has a linear relationship with the magnetic field.
Firstly, we prepare $|+\rangle
=\frac{1}{\sqrt{2}}(|0\rangle +|1\rangle )$. Secondly, let this state
\SCB{evolve under} the Hamiltonian for a time $t$, and we obtain $\frac{1}{\sqrt{2}}(e^{-i\frac{\omega}{2}t}|0\rangle +e^{i\frac{\omega}{2}t}|1\rangle )$.
Finally, we perform a projective measurement about $\hat{\sigma }_y$ on
this state, and we project this state into $|+_{y}\rangle
=\frac{1}{\sqrt{2}}(|0\rangle +i|1\rangle )$ with a probability of
$P_{+1}=\frac{1}{2}+\frac{1}{2}\sin \omega t$. Throughout of this paper,
we assume that the
magnetic field is weak ($\omega t\ll 1$), and so we have $P_{+1}\simeq
\frac{1}{2}+\frac{1}{2}\omega t$.
By repeating this
experiment many times, we can obtain the probability from the
experimental results and so the value of $\omega $ can be estimated.
The uncertainty of the estimation is given by
\begin{eqnarray}
 \delta \omega =\frac{\sqrt{P_{+1}(1-P_{+1})}}{|\frac{dP_{+1}}{d\omega }|}\frac{1}{\sqrt{N}}\label{sensitivity}
\end{eqnarray}
where $N$ denotes the number of repetitions of the experiments \cite{huelga1997improvement}.
We assume that the \SCB{interaction} time $t$ is much longer than the state
preparation time and measurement readout time. In this case, we have
$N\simeq \frac{T}{t}$ where $T$ is a given time for the sensing.
We can calculate the uncertainty as
\begin{eqnarray}
 \delta \omega \simeq \frac{1}{ \sqrt{Tt}}
\end{eqnarray}

We consider the magnetic field sensing under the effect of energy
relaxation.
The energy relaxation can be described by the standard Lindblad type
master equation as \cite{lindblad76,hornberger2009introduction}
\begin{eqnarray}
 &&\frac{d\rho (t)}{dt}=-i[H,\rho (t)]\nonumber \\
  &-&\frac{2\Gamma (1-s)}{2}(\sigma
  _+\sigma _-\rho(t)+\rho(t) \hat{\sigma }_+\sigma _- -2\sigma _-\rho (t)\sigma
   _+)\nonumber \\
 &-&\frac{2\Gamma s}{2}(\sigma
  _-\sigma _+\rho(t)
  +\rho(t) \hat{\sigma }_-\sigma _+ -2\sigma _+\rho (t)\sigma _-)
   \label{lindblad}
\end{eqnarray}
Here, $\Gamma $ denotes a decay rate while $s$ depends on the
temperature of the bath where $s=\frac{1}{2}$ ($s=0$) corresponds to an infinite
(a zero) temperature \cite{GZ01b}. An analytical solution for this master
equation is given as \cite{hein2005entanglement}
\begin{eqnarray}
\rho _I(t)=\frac{1}{4}(1+2e^{-\Gamma t}+e^{-2\Gamma
  t})\rho _0+\frac{1}{4}(1-e^{-2\Gamma t})\hat{\sigma }_x\rho _0\hat{\sigma
  }_x
  \nonumber \\
 + \frac{1}{4}(1-e^{-2\Gamma t})\hat{\sigma }_y\rho _0\hat{\sigma }_y+\frac{1}{4}(1-2e^{-\Gamma t}+e^{-2\Gamma
  t})\hat{\sigma }_z\rho_0\hat{\sigma
  }_z
  \nonumber \\
 +\frac{2s-1}{4}(1-e^{-2\Gamma t})(\hat{\sigma }_z\rho _0 +\rho _0\hat{\sigma
  }_z
  -i\hat{\sigma }_x\rho _0\hat{\sigma
  }_y+i\hat{\sigma }_y\rho _0\hat{\sigma
  }_x) \label{analytical}
\end{eqnarray}
where $\rho _I(t)=e^{iHt}\rho (t)e^{-iHt}$ and $\rho _0=\rho (0)$.
In the Ramsey type sequence with the energy relaxation, we obtain
\begin{eqnarray}
 &&P_{+1}={\rm{Tr}}[|+_{y}\rangle \langle +_y|\rho (t)]\nonumber \\
  &=&\frac{1}{2}+\frac{1}{2}e^{-\Gamma t}\sin \omega t\simeq
  \frac{1}{2}+\frac{1}{2}e^{-\Gamma t}\omega t. \label{ramseyp}
\end{eqnarray}
For the weak magnetic fields,
we can calculate the uncertainty from Eq. (\ref{sensitivity}) as
\begin{eqnarray}
 \delta \omega 
  \simeq \frac{2.33}{\sqrt{T/\Gamma }}
\end{eqnarray}
where we choose an optimum $t$
to minimize the uncertainty.

\section{Magnetic field sensing with quantum error correction}

We adopt a strategy to use the standard quantum error correction
technique for the magnetic
field sensing suggested in \cite{kessler2014quantum,arrad2014increasing,dur2014improved}. This requires two distinct qubits, namely a probe qubit and a memory qubit.
The probe qubit is
coupled with the magnetic field, while the interaction of the memory
qubit with the magnetic field is negligible.
On the other hand, the probe qubit is  affected by
energy relaxation while the memory qubit has a
much longer coherence time than the probe qubit.
Also, we assume that,  on these
two qubits, we can implement any unitary operations and
measurements with a much shorter time scale than the coherence time of
the qubits.

The Hamiltonian is described as
\begin{eqnarray}
 H=\frac{\omega }{2}\hat{\sigma }_z^{{\rm{(p)}}}+\frac{\omega '}{2}\hat{\sigma }_z^{{\rm{(m)}}}
\end{eqnarray}
where $\omega  $ ($\omega '$) denotes the resonant
frequency of the probe (memory) qubit. We assume
$\omega \gg \omega '$,
and so we use an approximation to
drop the term of
$\frac{\omega '}{2}\hat{\sigma}_z^{{\rm{(m)}}}$
from the Hamiltonian.
Also, the Lindblad master equation is described as
\begin{eqnarray}
&& \frac{d\rho (t)}{dt}=-i[H,\rho (t)]\nonumber \\
  &-&\frac{2\Gamma (1-s)}{2}(\sigma
  _+^{\rm{(p)}}\sigma ^{\rm{(p)}}_-\rho(t)+\rho(t) \hat{\sigma }^{\rm{(p)}}_+\sigma ^{\rm{(p)}}_- -2\sigma ^{\rm{(p)}}_-\rho (t)\sigma^{\rm{(p)}}
   _+)\nonumber \\
 &-&\frac{2\Gamma s}{2}(\sigma^{\rm{(p)}}
  _-\sigma ^{\rm{(p)}}_+\rho(t)
  +\rho(t) \hat{\sigma }^{\rm{(p)}}_-\sigma ^{\rm{(p)}}_+ -2\sigma ^{\rm{(p)}}_+\rho (t)\sigma ^{\rm{(p)}}_-)
  \ \  \label{lindbladp}
\end{eqnarray}
where only a probe qubit is affected by the energy relaxation.

\textcolor{black}{
\subsection{Parity measurement}
To demonstrate magnetic field sensing with quantum error correction
(QEC),
parity measurements are necessary to detect \SCB{bit flip errors}
\cite{kessler2014quantum,arrad2014increasing,dur2014improved,herrera2015quantum},
and we describe how we can construct the parity measurement by using the
two-qubit system without additional ancillary qubits. We define a controlled-not (CNOT) gate as follows
\begin{eqnarray}
 U_{\rm {CNOT}}=|0\rangle _{\rm {p}}\langle 0|\otimes \hat{\openone}_{\rm{m}}+|1\rangle _{\rm{p}}\langle
  1|\otimes \hat{\sigma }^{({\rm{m}})}_{x}
\end{eqnarray}
where the probe (memory) qubit is the control (target).
By performing the CNOT gate, a projective measurement in the basis of $\hat{\sigma
}_z^{({\rm{m}})}$, and another CNOT gate, we can construct the following
projective measurements.
\begin{eqnarray}
 \hat{\mathcal{P}}_{\rm{odd}}=U_{\rm {CNOT}}\frac{\hat{\openone}+\hat{\sigma }^{({\rm{m}})}_{z}}{2}U_{\rm {CNOT}}=|01\rangle _{\rm{pm}}\langle 01|+|10\rangle_{\rm{pm}}
\langle 10|\nonumber \\
 \hat{\mathcal{P}}_{\rm{even}}=U_{\rm {CNOT}}\frac{\hat{\openone}-\hat{\sigma }^{({\rm{m}})}_{z}}{2}U_{\rm {CNOT}}=|00\rangle _{\rm{pm}}\langle 00|+|11\rangle_{\rm{pm}}
\langle 11|\nonumber
\end{eqnarray}
and these are called parity measurements. }

\subsection{Single quantum error correction cycle}
By using this system, we can implement quantum field sensing as follows.
Let us consider a case when we implement the QEC cycle only one time.
Firstly, we prepare a state $\rho _0=|\phi _0\rangle _{\rm{pm}}\langle
\phi _0|$ where $|\phi_0\rangle _{\rm{pm}}=\frac{1}{\sqrt{2}}(|00\rangle _{{\rm
{pm}}}+|11\rangle _{{\rm {pm}}})$.
Secondly, let this state evolve by the Lindblad master equation in
Eq. (\ref{lindbladp}) for a time $t$.
 Thirdly, we perform a parity projection
 to check if a bit-flip error occurs on the probe qubit.
If the parity is even (odd), a projective operator described by $\hat{\mathcal{P}}_{{\rm{even}}}=|00\rangle _{\rm{pm}}\langle 00|+|11\rangle_{\rm{pm}}
\langle 11|$ 
($\hat{\mathcal{P}}_{{\rm{odd}}}=|01\rangle _{\rm{pm}}\langle 01|+|10\rangle_{\rm{pm}}
\langle 10|$) is applied on the quantum states.
If the parity is odd, this provides us with the existence of the
bit-flip error on the probe qubit, and so we will apply a
 recovery operation $\hat{\sigma
}_x^{(\rm{p})}$ on the probe qubit.
Finally, we measure the state in the basis of
$|\psi ^{(\pm)}_{\rm{f}}\rangle _{{\rm{pm}}}=\frac{1}{\sqrt{2}}(|00\rangle _{{\rm {pm}}}\pm i|11\rangle _{{\rm{pm}}})$ where the projection
operator is described as $\hat{\mathcal{P}}^{{\rm {(pm)}}}_{{{\rm {f}}},\pm
}=|\psi ^{(\pm )}_{\rm{f}}\rangle _{{\rm{pm}}} \langle \psi ^{(\pm )}_{\rm{f}}|$.
In this case,
the readout probability can be calculated as
\begin{eqnarray}
 P^{({\rm {QEC}})}_{+1}={\rm{Tr}}[\hat{\mathcal{P}}^{{\rm {(pm)}}}_{{{\rm {f}}},+}\rho^{{\rm {(QEC)}}}
  (t)]
  =\frac{1}{2}+\frac{1}{2}e^{-\Gamma t}\sin \omega t.
\end{eqnarray}
Interestingly, this is the same probability as the Eq. (\ref{ramseyp}),
and so the uncertainty of the estimation with QEC is the same as
that of the \textcolor{black}{single-qubit sensors}.

\subsection{Multiple quantum error correction cycle}

\begin{figure}[h] 
\begin{center}
\includegraphics[width=0.9\linewidth]{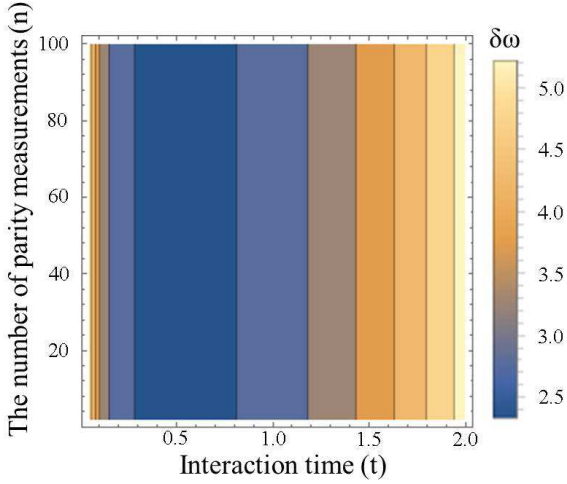} 
\caption{Uncertainty of the estimation of the magnetic fields with
 multiple QEC cycles under the effect of energy relaxation. The parameters are chosen as $s=0.5$ and $\Gamma =1$. We use two qubits to construct an encoded state,
 and one of the qubits is affected by energy relaxation.
 By performing parity measurements with a time period of $\frac{t}{n}$
 where $t$ denotes a total evolution time and $n$ denotes the number of the parity
 measurements, we can detect the bit flip error,
 and the subsequent bit-flip operation can recover the state. However,
 these numerical results show that
 the uncertainty does not depend on $n$, and so we cannot improve the
 sensitivity even if we increase the number of QEC cycles.
 The minimum uncertainty is still 2.33, as with the single-qubit sensor.
 }
 \label{qec}
\end{center}
\end{figure}

Let us consider a case when we implement multiple QEC cycles.
The first step is to prepare a state $\frac{1}{\sqrt{2}}(|00\rangle _{{\rm
{pm}}}+|11\rangle _{{\rm {pm}}})$.
The second step is a time evolution of the system by the Lindblad master equation in
Eq. (\ref{lindblad}) for a time $\tau =t/n$ where $n$ denotes the number of
the parity measurements.
The third step is to perform a parity projection.
If the parity is odd,  we will apply the recovery operation $\hat{\sigma
}_x^{(\rm{p})}$ on the probe qubit.
The final step is that, after repeating the second step and the third step $n$ times,
 we readout the state via the measurement in the basis of
$\frac{1}{\sqrt{2}}(|00\rangle _{{\rm {pm}}}\pm i|11\rangle _{{\rm{pm}}})$ where the projective
operator is described as $\hat{\mathcal{P}}^{{\rm {(pm)}}}_{\pm
1}=\frac{1}{2}(|00\rangle _{{\rm {pm}}}\pm i|11\rangle _{{\rm{pm}}})
(\langle 00|\pm i{} _{{\rm{pm}}}\langle 11|)$.
We numerically calculate the uncertainty $\delta \omega $ for this
protocol, and plot it in the Fig \ref{qec}.
Interestingly, $\delta \omega$ does not depend on $n$, and the
implementation of the multiple QEC does not improve the sensitivity over
the standard Ramsey type scheme.
Although we plot the case of $s=0.5$ in the Fig \ref{qec}, we obtain the same results
 for other values of $s$.

 Let us discuss a possible reason why QEC does not improve the
sensitivity. For simplicity, we \SCB{specifically} discuss the case for $s=0.5$.
From  Eq. (\ref{analytical}), we know that the qubit is affected by three types of
Pauli noise, namely $\hat{\sigma }_x$, $\hat{\sigma }_y$, and
$\hat{\sigma }_z$.
Importantly, since the initial state is an eigenstate of $\hat{\sigma
}_x$, \SCB{consequently it is} $\hat{\sigma }_y$ and
$\hat{\sigma }_z$ errors mainly decohere the qubit.
On the other hand, if we use the two-qubit encoded state, \SCB{then}
$\hat{\sigma }^{(\rm{p})}_x$, $\hat{\sigma }^{(\rm{p})}_y$, and $\hat{\sigma }^{(\rm{p})}_z$
\SCB{all induce} decoherence. Moreover, QEC with the two-qubit code
cannot distinguish the error of $\hat{\sigma }^{(\rm{p})}_x$ from the
error of $\hat{\sigma }^{(\rm{p})}_y$, and so the recovery operation can
remove only one of these
two errors. Therefore, even if we implement QEC, two types of error still
remains in the quantum states, which is effectively the same as the case
of the single-qubit strategy. This qualitatively explains why QEC cannot
improve the sensitivity for the metrology under the effect of energy relaxation.

\section{Magnetic field sensing with quantum error detection}
Here, we propose to use a postselection where we discard the
state when we detect the bit-flip error, \SCB{which we refer to} as a quantum
error detection (QED) \SCB{strategy}.
\SCB{As with the} QEC strategy, we will use two different systems, namely a
probe qubit and a memory qubit so that we can encode the state in a
logical basis.
Surprisingly, we will show that this postselection actually provides us
with quantum enhancement over the \textcolor{black}{single-qubit sensors}.

\subsection{Single quantum error detection}
Let us discuss the case where we implement the QED only one time.
In QED strategy, we use the same sequence as QEC except that we 
discard the state when the error is detected by the parity projection.
The final state before the measurement readout is
calculated as
\begin{eqnarray}
\rho ^{{\rm{(QED)}}}(t)=\frac{1}{(\frac{1+e^{-2\Gamma t}}{2})}e^{-iHt}\Big{(}\frac{1}{4}(1+2e^{-\Gamma t}+e^{-2\Gamma
  t})\rho _0\nonumber \\
  +\frac{1}{4}(1-2e^{-\Gamma t}+e^{-2\Gamma
  t})\hat{\sigma }^{(\rm{p})}_z\rho _0\hat{\sigma
  }^{(\rm{p})}_z\nonumber \\
 +\frac{2s-1}{4}(1-e^{-2\Gamma t})(\hat{\sigma }^{(\rm{p})}_z\rho _0 +\rho_0\hat{\sigma
  }^{(\rm{p})}_z)
  \Big{)}e^{iHt}
  \nonumber
\end{eqnarray}
where the success probability to obtain this state is
$\frac{1+e^{-2\Gamma t}}{2}$.
The probability to project this state into $\hat{\mathcal{P}}^{{\rm
{(pm)}}}_{{{\rm {f}}},+}$ is calculated as
\begin{eqnarray}
 P^{({\rm {QED}})}_{+1}={\rm{Tr}}[\hat{\mathcal{P}}^{{\rm {(pm)}}}_{{{\rm {f}}},+}\rho^{{\rm {(QED)}}}
  (t)]
  =\frac{1}{2}+\frac{1}{2}\frac{\sin \omega t}{(\frac{e^{\Gamma t}+e^{-\Gamma t}}{2})}
\end{eqnarray}
and so the uncertainty is given as
 \begin{eqnarray}
 \delta \omega \simeq \frac{e^{\Gamma t}+e^{-\Gamma
  t}}{2t}\frac{1}{\sqrt{N}}=\sqrt{\frac{e^{2\Gamma t}+1}{2}}\frac{1}{\sqrt{Tt}}
 \end{eqnarray}
 where $N\simeq \sqrt{\frac{T}{t}\frac{1+e^{-2\Gamma t}}{2}} $ denotes
 the number of the measurement readouts
 when no error is detected.
 For weak magnetic fields, we obtain
 \begin{eqnarray}
  \delta \omega \simeq \frac{1.895}{\sqrt{T/\Gamma }}
 \end{eqnarray}
 where we choose $t$
 to minimize $\delta \omega$,
 and this show that the QED strategy is better than the \textcolor{black}{single-qubit sensors}.

 Let us discuss a possible reason why the QED can improve the
sensitivity. 
The probe qubit is affected by three type of
Pauli noise, namely $\hat{\sigma }^{(\rm{p})}_x$, $\hat{\sigma }^{(\rm{p})}_y$, and
$\hat{\sigma }^{(\rm{p})}_z$. If $\hat{\sigma }^{(\rm{p})}_x$ or
$\hat{\sigma }^{(\rm{p})}_y$ is applied, we can detect this error, and
the state can be discarded. In our strategy, only dephasing (an error
defined by $\hat{\sigma }_z$) is relevant
to decrease the sensitivity, and this makes our scheme better than the
\textcolor{black}{single-qubit scheme}
where both $\hat{\sigma }^{(\rm{p})}_y$ and
$\hat{\sigma }^{(\rm{p})}_z$ decrease the coherence of the state.

\subsection{Multiple quantum error detection}
We discuss the case when we implement multiple QED \SCB{rounds}.
We use the same sequence as the multiple QEC, and we perform parity
projection $n$ times before the readout. The only difference \SCB{is that we now merely postselect rather than correcting}.
\SCB{At the end, we} readout the state
only if we do not detect any errors within the $n$ parity
measurements. Otherwise, we will discard the state before the readout measurement.

  \begin{figure}[h] 
\begin{center}
\includegraphics[width=0.9\linewidth]{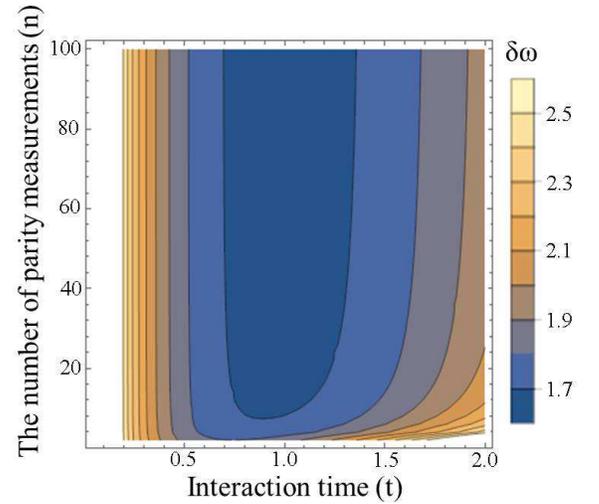} 
\caption{Uncertainty of the estimation of the magnetic fields with
 multiple QED cycles for $s=0.5$ and $\Gamma =1$. We use the same sequence as \SCB{for }the QEC except that we
 discard the state before the readout if we detect any errors in the
 parity projections. As we increase the number of the parity
 measurements, we can decrease the uncertainty, and the minimum
 uncertainty is $\delta \omega \simeq 1.65
  /\sqrt{T\Gamma }$, which beats the one in \textcolor{black}{the single-qubit sensors}.
 }
 \label{qed}
\end{center}
\end{figure}

  \begin{figure}[h] 
\begin{center}
\includegraphics[width=0.9\linewidth]{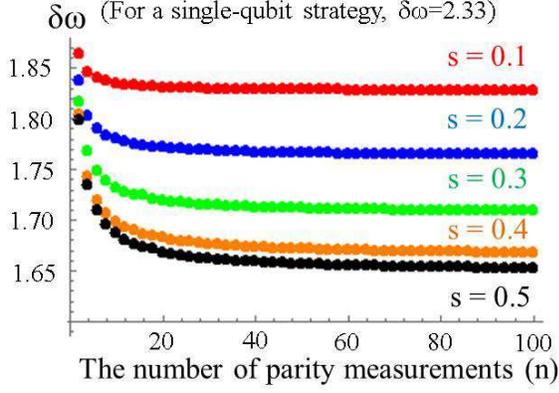} 
\caption{Uncertainty of the estimation of the magnetic fields with
 multiple QED cycles  against the number of the parity
 measurements $n$. Here, we choose the interaction
 time $t$ to minimize the uncertainty, and we fix $\Gamma =1$. 
 }
 \label{2qed}
\end{center}
\end{figure}

We consider a case $s=\frac{1}{2}$ where we can get an analytical
solution for the uncertainty of the estimation. Also, for simplicity, we assume $n$ is an even number.
The state before the readout is calculated as
\begin{eqnarray}
\rho ^{(\rm{MQED})}_n(t )
 =e^{-iHt}\Big{(}
 \big{(}\sum_{m=0}^{\frac{n}{2}}{}_nC_{2m}(p_1)^{n-2m}(1-p_1)^{2m}
  \big{)}\rho _0\nonumber \\
  +\big{(}\sum_{m=1}^{\frac{n}{2}}{}_nC_{2m-1}(p_1)^{n-2m+1}(1-p_1)^{2m-1}
  \big{)}\hat{\sigma }_z^{(1)}\rho _0\hat{\sigma }_z^{(1)}\Big{)}e^{iHt}
  \nonumber 
\end{eqnarray}
where $p_1=\frac{\frac{1}{4}(1+2e^{-\Gamma \tau }+e^{-2\Gamma
  \tau})}{(\frac{1+e^{-2\Gamma \tau }}{2})} $ and ${}_nC_{m}=\frac{n!}{(n-m)!m!}$. 
 We obtain this state with a success
probability of $(\frac{1+e^{-2\Gamma \tau }}{2})^n$.
We consider the following probability.
\begin{eqnarray}
 P^{(\rm{MQED})}_{+1}={\rm{Tr}}[|+^{(y)}_{\rm{L}}\rangle \langle
  +^{(y)}_{\rm{L}}|\rho ^{(\rm{MQED})}_n(t )]\ \ \ 
  \nonumber \\
 =\Big{(}\sum_{m=0}^{\frac{n}{2}}{}_nC_{2m}(p_1)^{n-2m}(1-p_1)^{2m}
  \Big{)}
  \frac{1+\sin \omega t}{2}\nonumber \\
  +\Big{(}\sum_{m=1}^{\frac{n}{2}}{}_nC_{2m-1}(p_1)^{n-2m+1}(1-p_1)^{2m-1}
  \Big{)} \frac{1-\sin \omega t}{2}\nonumber \\
\end{eqnarray}
where $|+^{(y)}_{\rm{L}}\rangle =\frac{1}{\sqrt{2}}(|00\rangle
+i|11\rangle) $.
The sensitivity is given as
\begin{eqnarray}
 \delta \omega &=&
  \frac{\sqrt{P^{(\rm{MQED})}_{+1}(1-P^{(\rm{MQED})}_{+1})}}{|\frac{dP^{(\rm{MQED})}_{+1}}{d\omega }|}\cdot
  \frac{1}{\sqrt{\frac{T}{t}(\frac{1+e^{-2\Gamma
  \frac{t}{n}}}{2})^n}}\nonumber \\
  &\simeq& 1.65
  /\sqrt{T\Gamma } \label{mqed}
\end{eqnarray}
where we choose $t$ and $n$ to minimize $\delta \omega$.
We plot the $\delta \omega $ to show the dependence of $t$ and $n$ in
the Fig. \ref{qed}. As we increase the number of the parity
measurements, the uncertainty increases and converges to the value
described in
the Eq. (\ref{mqed}).

We also perform numerical simulations to calculate the uncertainty of
the estimation in the QED strategy for other values of $s$. The results
are plotted in the Fig. \ref{2qed}. We confirmed that the QED strategy
actually beats the \textcolor{black}{single-qubit sensors}
 for the other values of $s$.

 \textcolor{black}{We discuss an intuitive reason why multiple QED strategy
 can beat
 the single QED strategy. If we perform a single parity measurement in
 the end of the time evolution, there will be
 a possibility that bit flip errors occur twice within the time
 evolution, which induces undetected error. Multiple QED provides us
 with a capability to eliminate such a possibility so that we can
 significantly suppress the effect of the bit flip error.}

\subsection{Adaptive feedback}

\begin{figure}[h] 
\begin{center}
\includegraphics[width=1.0\linewidth]{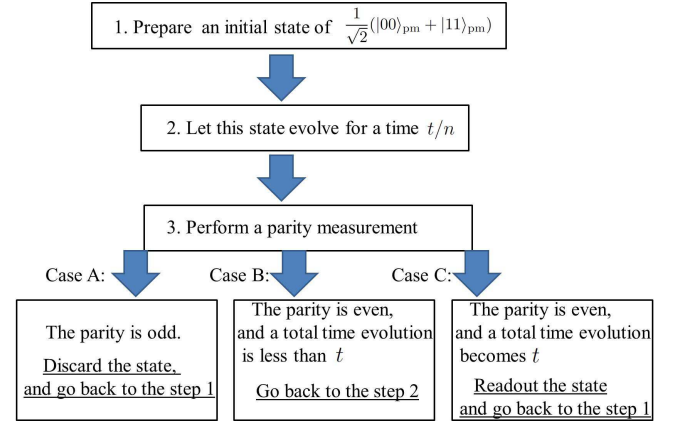} 
\caption{The schematic to perform an adaptive feedback in our QED
 strategy to estimate the target magnetic fields. Depending on the
 measurement results, we will implement different operations.
 }
 \label{adaptive}
\end{center}
\end{figure}

Interestingly, we can further improve the sensitivity by using an
adaptive feedback.
In the last subsection, we discussed multiple QED rounds
 where we use the same sequence as QEC 
except that we
 discard the state before the readout if we detect any errors in the
 parity projections. However, this strategy is inefficient, because we
 waste time between the parity measurement and the readout \SCB{once we have detected an}
 error. For example, when we detect the error at $k$-th parity measurements,
 we have a time $\frac{n-k}{n}t$ before the readout, and we spend this
 time without \SCB{contributing to the sensitivity}.
 To improve this point, we propose to use an adaptive feedback where we immediately initialize the
 state for the next round
 whenever we detect the error, as shown in the Fig. \ref{adaptive}.

   \begin{figure}[h] 
\begin{center}
\includegraphics[width=1.0\linewidth]{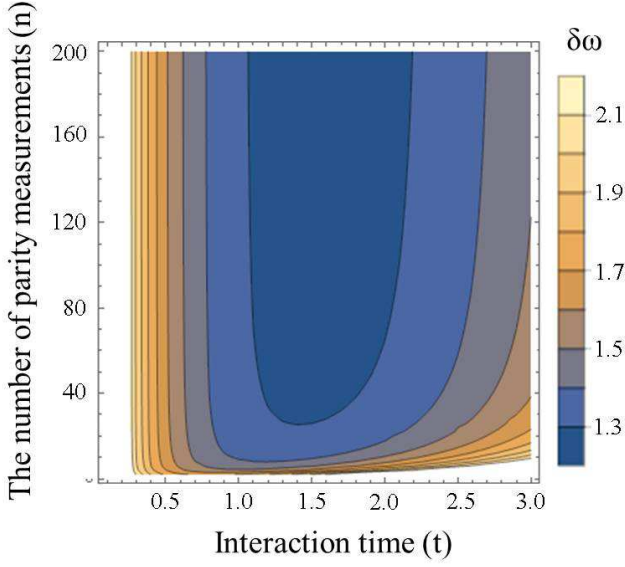} 
\caption{Uncertainty of the estimation of the magnetic fields with
 adaptive feedback for $s=0.5$ and $\Gamma
 =1$. The detail of the procedure is given in the Fig. \ref{adaptive}.
 As we increase the number of the parity
 measurements, we can decrease the uncertainty, and the minimum
 uncertainty is $\delta \omega \simeq 1.25
  /\sqrt{T\Gamma }$, which is better than the QED strategy without adaptive feedback.
 }
 \label{simonqed}
\end{center}
\end{figure}

 Firstly, we consider a case for $s=0.5$ to obtain an analytical solution.
 In this adaptive feedback strategy, we can calculate the average time
 for a single cycle of the sensing as follows. If we detect the error in the
 $k$ the parity measurements, the time between the state preparation and the
 last parity projection is $\frac{k}{n}t$. This event occurs with a
 probability of $(p_s)^{k-1}(1-p_s)$ where $p_s=\frac{1+e^{-2\Gamma \tau }}{2}$
 denotes a probability that we do not detect the error. On the other
 hand, if we do not detect any errors for $n$ parity measurements (which
 occurs with a probability of $(p_s)^n$), the
 time between the state preparation and the final readout is $t$. So the average time for a
 single cycle of this adaptive feedback strategy is
\begin{eqnarray}
 t_{\rm{av}}=(p_s)^nt+\sum_{k=1}^{n}(p_s)^{k-1}(1-p_s)k\frac{t}{n}.
\end{eqnarray}
We can calculate the number of the readout measurements for a
given time $T$ as $N=\frac{T}{t_{\rm{av}}}(\frac{1+e^{-2\Gamma
  \frac{t}{n}}}{2})^n$.
The sensitivity is given as
\begin{eqnarray}
 \delta \omega &=&
  \frac{\sqrt{P^{(\rm{MQED})}_{+1}(1-P^{(\rm{MQED})}_{+1})}}{|\frac{dP^{(\rm{MQED})}_{+1}}{d\omega }|}\cdot
  \frac{1}{\sqrt{\frac{T}{t_{\rm{av}}}(\frac{1+e^{-2\Gamma
  \frac{t}{n}}}{2})^n}}\nonumber \\
  &\simeq& 1.25
  /\sqrt{T\Gamma } \label{amqed}
\end{eqnarray}
where we choose $t$ and $n$ to minimize $\delta \omega$.
We plot the $\delta \omega $ to show the dependence of $t$ and $n$ in
the Fig. \ref{simonqed}. Again, as we increase the number of the parity
measurements, the uncertainty increases and converges to the value of
the Eq. (\ref{amqed}).
 
  \begin{figure}[h] 
\begin{center}
\includegraphics[width=0.9\linewidth]{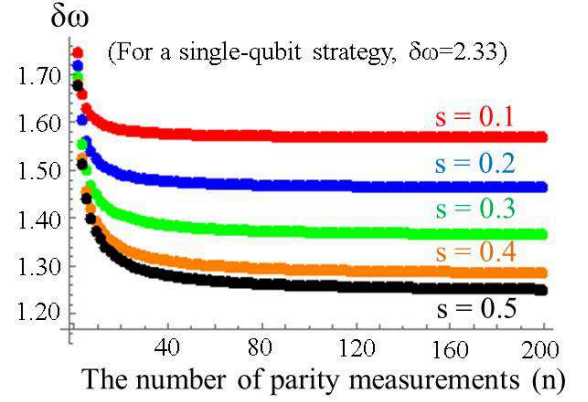} 
\caption{Uncertainty of the estimation of the magnetic fields with
 adaptive feedback  against the number of the parity
 measurements $n$. Here, we choose the interaction
 time $t$ to minimize the uncertainty, and we fix $\Gamma =1$. 
 }
 \label{3qed}
\end{center}
\end{figure}

We also performed numerical simulations to calculate the uncertainty of
the estimation in the adaptive feedback strategy for other values of $s$. 
\SCB{The results are plotted in Fig. \ref{3qed}}. We confirmed that the adaptive strategy improves
the sensitivity over the non-adaptive strategy.

\subsection{Imperfect parity projection}
    \begin{figure}[h] 
\begin{center}
\includegraphics[width=0.9\linewidth]{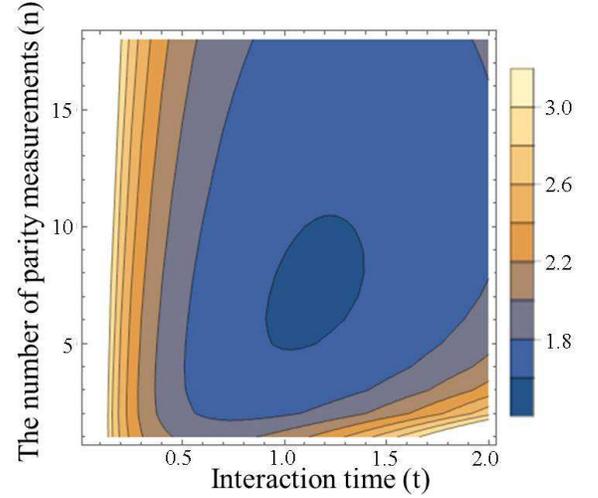} 
\caption{Uncertainty of the estimation of the magnetic fields by using
 imperfect parity projections with a finite error rate of
 $\epsilon$. Here, we fix $s=0.5$, $\epsilon =0.02$, and $\Gamma =1$.
 There exist an optimal set of the interaction time and the number of measurements.
 }
 \label{threeerror}
\end{center}
\end{figure}
   \begin{figure}[h] 
\begin{center}
\includegraphics[width=0.9\linewidth]{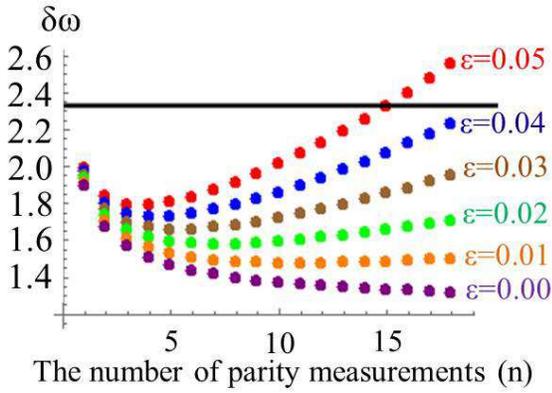} 
\caption{Uncertainty of the estimation of the magnetic fields by using
 imperfect parity projections with a finite error rate of $\epsilon$. Here, we choose the interaction
 time $t$ to minimize the uncertainty, and we fix $s=0.5$ and $\Gamma =1$.  The solid
 line denotes the case of the \textcolor{black}{single-qubit sensors}. By using the imperfect parity
 measurement with an error rate of $\epsilon$, our QED strategy with
 adaptive feedback can beat \textcolor{black}{single-qubit sensors}.
 }
 \label{error}
\end{center}
\end{figure}

  \begin{figure}[h] 
\begin{center}
\includegraphics[width=0.9\linewidth]{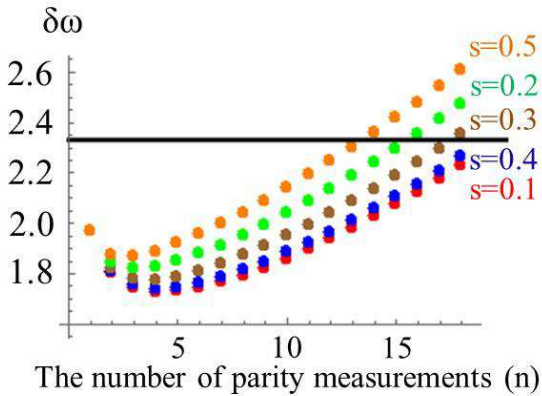} 
\caption{Uncertainty of the estimation of the magnetic fields by using
 imperfect parity projections with a finite error rate of $\epsilon$. Here, we choose the interaction
 time $t$ to minimize the uncertainty, and we fix $\epsilon=0.04$ and $\Gamma =1$.  The solid
 line denotes the case of \textcolor{black}{single-qubit sensors}.
 }
 \label{serror}
\end{center}
\end{figure}
\SCB{Finally} we consider an effect of imperfect parity projections. In the QED
strategy, as we increase the number of the parity projections, the
uncertainty of the estimation decreases, if we assume ideal parity
projections.
In the actual experiments, we cannot avoid \SCB{the possibility of} an error in the parity
measurement, \SCB{and thus the optimal number of parity measurements will be a finite.} 
We consider a model that depolarized noise occurs with a probability of $\epsilon$.
 If the measurement result is even, we obtain
  \begin{eqnarray}
  \rho '=  (1-\epsilon )
\frac{   \hat{\mathcal{P}}_{\rm{even}}. \rho .
\hat{\mathcal{P}}_{\rm{even}}}{{\rm{Tr}}[
\hat{\mathcal{P}}_{\rm{even}}. \rho 
\hat{\mathcal{P}}_{\rm{even}}]}
+\epsilon \frac{1}{4}\hat{\openone}_{\rm{pm}}
 \end{eqnarray}
where the state becomes a completely mixed state with a probability of
$\epsilon$.
With this noise model, we numerically evaluate the performance of the adaptive feedback
 strategy with imperfect parity measurements.
As the Figs \ref{threeerror}, \ref{error}, and \ref{serror} show, we can
beat \textcolor{black}{the single-qubit sensors}
as long as the error rate $\epsilon$ is around \SCB{four} percent.

\textcolor{black}{We discuss a possible reason why the large error of
around \SCB{four} percent can be tolerated in our scheme. If \SCB{depolarising} 
noise occurs  on a qubit, this
can be interpreted as a random applications of Pauli matrices $\hat{\openone}$,
$\hat{\sigma }_x$, $\hat{\sigma }_y$, and $\hat{\sigma }_z$ with equally
probability. However, the parity measurement in the next round can
eliminate the degrading effect of $\hat{\sigma }_x$ and $\hat{\sigma
}_y$. So \SCB{only one of the three noise operators can fully impact}
the sensitivity of our scheme. }

\section{Conclusion}
In conclusion, we have investigated \SCB{the} performance of quantum error correction
to improve quantum field sensing under the effect of dephasing.
We have shown that the standard quantum error correction, including error
detection and recovery operations, does not
improve the sensitivity over the \textcolor{black}{single-qubit sensors}.
However, we have found that, if we adopt a postselection to discard the state whenever an error
is detected, we can actually achieve \SCB{significant} quantum enhancement, \SCB{even when the operations we use are imperfect.}
Since  energy relaxation is one of the typical noise \SCB{types} in solid state
systems, our results pave a way to realize a quantum enhanced sensors in
realistic conditions.

This work was supported by JSPS KAKENHI Grants
15K17732.
Also, this work was partly supported by MEXT KAKENHI
 Grant Number 15H05870.

\begin{thebibliography}{30}
\expandafter\ifx\csname natexlab\endcsname\relax\def\natexlab#1{#1}\fi
\expandafter\ifx\csname bibnamefont\endcsname\relax
  \def\bibnamefont#1{#1}\fi
\expandafter\ifx\csname bibfnamefont\endcsname\relax
  \def\bibfnamefont#1{#1}\fi
\expandafter\ifx\csname citenamefont\endcsname\relax
  \def\citenamefont#1{#1}\fi
\expandafter\ifx\csname url\endcsname\relax
  \def\url#1{\texttt{#1}}\fi
\expandafter\ifx\csname urlprefix\endcsname\relax\def\urlprefix{URL }\fi
\providecommand{\bibinfo}[2]{#2}
\providecommand{\eprint}[2][]{\url{#2}}

\bibitem[{\citenamefont{Simon}(1999)}]{simon1999local}
\bibinfo{author}{\bibfnamefont{J.}~\bibnamefont{Simon}},
  \bibinfo{journal}{Advances in Physics} \textbf{\bibinfo{volume}{48}},
  \bibinfo{pages}{449} (\bibinfo{year}{1999}).

\bibitem[{\citenamefont{Chang et~al.}(1992)\citenamefont{Chang, Hallen,
  Harriott, Hess, Kao, Kwo, Miller, Wolfe, Van~der Ziel, and
  Chang}}]{chang1992scanning}
\bibinfo{author}{\bibfnamefont{A.}~\bibnamefont{Chang}},
  \bibinfo{author}{\bibfnamefont{H.}~\bibnamefont{Hallen}},
  \bibinfo{author}{\bibfnamefont{L.}~\bibnamefont{Harriott}},
  \bibinfo{author}{\bibfnamefont{H.}~\bibnamefont{Hess}},
  \bibinfo{author}{\bibfnamefont{H.}~\bibnamefont{Kao}},
  \bibinfo{author}{\bibfnamefont{J.}~\bibnamefont{Kwo}},
  \bibinfo{author}{\bibfnamefont{R.}~\bibnamefont{Miller}},
  \bibinfo{author}{\bibfnamefont{R.}~\bibnamefont{Wolfe}},
  \bibinfo{author}{\bibfnamefont{J.}~\bibnamefont{Van~der Ziel}},
  \bibnamefont{and} \bibinfo{author}{\bibfnamefont{T.}~\bibnamefont{Chang}},
  \bibinfo{journal}{Applied physics letters} \textbf{\bibinfo{volume}{61}},
  \bibinfo{pages}{1974} (\bibinfo{year}{1992}).

\bibitem[{\citenamefont{Poggio and Degen}(2010)}]{poggio2010force}
\bibinfo{author}{\bibfnamefont{M.}~\bibnamefont{Poggio}} \bibnamefont{and}
  \bibinfo{author}{\bibfnamefont{C.}~\bibnamefont{Degen}},
  \bibinfo{journal}{Nanotechnology} \textbf{\bibinfo{volume}{21}},
  \bibinfo{pages}{342001} (\bibinfo{year}{2010}).

\bibitem[{\citenamefont{Budker and Romalis}(2007)}]{budker2007optical}
\bibinfo{author}{\bibfnamefont{D.}~\bibnamefont{Budker}} \bibnamefont{and}
  \bibinfo{author}{\bibfnamefont{M.}~\bibnamefont{Romalis}},
  \bibinfo{journal}{Nature Physics} \textbf{\bibinfo{volume}{3}},
  \bibinfo{pages}{227} (\bibinfo{year}{2007}).

\bibitem[{\citenamefont{Maze et~al.}(2008)\citenamefont{Maze, Stanwix, Hodges,
  Hong, Taylor, Cappellaro, Jiang, Dutt, Togan, Zibrov
  et~al.}}]{maze2008nanoscale}
\bibinfo{author}{\bibfnamefont{J.}~\bibnamefont{Maze}},
  \bibinfo{author}{\bibfnamefont{P.}~\bibnamefont{Stanwix}},
  \bibinfo{author}{\bibfnamefont{J.}~\bibnamefont{Hodges}},
  \bibinfo{author}{\bibfnamefont{S.}~\bibnamefont{Hong}},
  \bibinfo{author}{\bibfnamefont{J.}~\bibnamefont{Taylor}},
  \bibinfo{author}{\bibfnamefont{P.}~\bibnamefont{Cappellaro}},
  \bibinfo{author}{\bibfnamefont{L.}~\bibnamefont{Jiang}},
  \bibinfo{author}{\bibfnamefont{M.}~\bibnamefont{Dutt}},
  \bibinfo{author}{\bibfnamefont{E.}~\bibnamefont{Togan}},
  \bibinfo{author}{\bibfnamefont{A.}~\bibnamefont{Zibrov}},
  \bibnamefont{et~al.}, \bibinfo{journal}{Nature}
  \textbf{\bibinfo{volume}{455}}, \bibinfo{pages}{644} (\bibinfo{year}{2008}),
  ISSN \bibinfo{issn}{0028-0836}.

\bibitem[{\citenamefont{Taylor et~al.}(2008)\citenamefont{Taylor, Cappellaro,
  Childress, Jiang, Budker, Hemmer, Yacoby, Walsworth, and
  Lukin}}]{taylor2008high}
\bibinfo{author}{\bibfnamefont{J.}~\bibnamefont{Taylor}},
  \bibinfo{author}{\bibfnamefont{P.}~\bibnamefont{Cappellaro}},
  \bibinfo{author}{\bibfnamefont{L.}~\bibnamefont{Childress}},
  \bibinfo{author}{\bibfnamefont{L.}~\bibnamefont{Jiang}},
  \bibinfo{author}{\bibfnamefont{D.}~\bibnamefont{Budker}},
  \bibinfo{author}{\bibfnamefont{P.}~\bibnamefont{Hemmer}},
  \bibinfo{author}{\bibfnamefont{A.}~\bibnamefont{Yacoby}},
  \bibinfo{author}{\bibfnamefont{R.}~\bibnamefont{Walsworth}},
  \bibnamefont{and} \bibinfo{author}{\bibfnamefont{M.}~\bibnamefont{Lukin}},
  \bibinfo{journal}{Nature Physics} \textbf{\bibinfo{volume}{4}},
  \bibinfo{pages}{810} (\bibinfo{year}{2008}).

\bibitem[{\citenamefont{Balasubramanian
  et~al.}(2008)\citenamefont{Balasubramanian, Chan, Kolesov, Al-Hmoud, Tisler,
  Shin, Kim, Wojcik, Hemmer, Krueger et~al.}}]{balasubramanian2008nanoscale}
\bibinfo{author}{\bibfnamefont{G.}~\bibnamefont{Balasubramanian}},
  \bibinfo{author}{\bibfnamefont{I.}~\bibnamefont{Chan}},
  \bibinfo{author}{\bibfnamefont{R.}~\bibnamefont{Kolesov}},
  \bibinfo{author}{\bibfnamefont{M.}~\bibnamefont{Al-Hmoud}},
  \bibinfo{author}{\bibfnamefont{J.}~\bibnamefont{Tisler}},
  \bibinfo{author}{\bibfnamefont{C.}~\bibnamefont{Shin}},
  \bibinfo{author}{\bibfnamefont{C.}~\bibnamefont{Kim}},
  \bibinfo{author}{\bibfnamefont{A.}~\bibnamefont{Wojcik}},
  \bibinfo{author}{\bibfnamefont{P.}~\bibnamefont{Hemmer}},
  \bibinfo{author}{\bibfnamefont{A.}~\bibnamefont{Krueger}},
  \bibnamefont{et~al.}, \bibinfo{journal}{Nature}
  \textbf{\bibinfo{volume}{455}}, \bibinfo{pages}{648} (\bibinfo{year}{2008}).

\bibitem[{\citenamefont{Huelga et~al.}(1997)\citenamefont{Huelga, Macchiavello,
  Pellizzari, Ekert, Plenio, and Cirac}}]{huelga1997improvement}
\bibinfo{author}{\bibfnamefont{S.}~\bibnamefont{Huelga}},
  \bibinfo{author}{\bibfnamefont{C.}~\bibnamefont{Macchiavello}},
  \bibinfo{author}{\bibfnamefont{T.}~\bibnamefont{Pellizzari}},
  \bibinfo{author}{\bibfnamefont{A.}~\bibnamefont{Ekert}},
  \bibinfo{author}{\bibfnamefont{M.}~\bibnamefont{Plenio}}, \bibnamefont{and}
  \bibinfo{author}{\bibfnamefont{J.}~\bibnamefont{Cirac}},
  \bibinfo{journal}{Phys. Rev. Lett.} \textbf{\bibinfo{volume}{79}},
  \bibinfo{pages}{3865} (\bibinfo{year}{1997}).

\bibitem[{\citenamefont{Degen et~al.}(2016)\citenamefont{Degen, Reinhard, and
  Cappellaro}}]{degen2016quantum}
\bibinfo{author}{\bibfnamefont{C.}~\bibnamefont{Degen}},
  \bibinfo{author}{\bibfnamefont{F.}~\bibnamefont{Reinhard}}, \bibnamefont{and}
  \bibinfo{author}{\bibfnamefont{P.}~\bibnamefont{Cappellaro}},
  \bibinfo{journal}{arXiv preprint arXiv:1611.02427}  (\bibinfo{year}{2016}).

\bibitem[{\citenamefont{Kessler et~al.}(2014)\citenamefont{Kessler, Lovchinsky,
  Sushkov, and Lukin}}]{kessler2014quantum}
\bibinfo{author}{\bibfnamefont{E.~M.} \bibnamefont{Kessler}},
  \bibinfo{author}{\bibfnamefont{I.}~\bibnamefont{Lovchinsky}},
  \bibinfo{author}{\bibfnamefont{A.~O.} \bibnamefont{Sushkov}},
  \bibnamefont{and} \bibinfo{author}{\bibfnamefont{M.~D.} \bibnamefont{Lukin}},
  \bibinfo{journal}{Phys. Rev. Lett.} \textbf{\bibinfo{volume}{112}},
  \bibinfo{pages}{150802} (\bibinfo{year}{2014}).

\bibitem[{\citenamefont{Arrad et~al.}(2014)\citenamefont{Arrad, Vinkler,
  Aharonov, and Retzker}}]{arrad2014increasing}
\bibinfo{author}{\bibfnamefont{G.}~\bibnamefont{Arrad}},
  \bibinfo{author}{\bibfnamefont{Y.}~\bibnamefont{Vinkler}},
  \bibinfo{author}{\bibfnamefont{D.}~\bibnamefont{Aharonov}}, \bibnamefont{and}
  \bibinfo{author}{\bibfnamefont{A.}~\bibnamefont{Retzker}},
  \bibinfo{journal}{Phys. Rev. Lett.} \textbf{\bibinfo{volume}{112}},
  \bibinfo{pages}{150801} (\bibinfo{year}{2014}).

\bibitem[{\citenamefont{D{\"u}r et~al.}(2014)\citenamefont{D{\"u}r,
  Skotiniotis, Froewis, and Kraus}}]{dur2014improved}
\bibinfo{author}{\bibfnamefont{W.}~\bibnamefont{D{\"u}r}},
  \bibinfo{author}{\bibfnamefont{M.}~\bibnamefont{Skotiniotis}},
  \bibinfo{author}{\bibfnamefont{F.}~\bibnamefont{Froewis}}, \bibnamefont{and}
  \bibinfo{author}{\bibfnamefont{B.}~\bibnamefont{Kraus}},
  \bibinfo{journal}{Phys. Rev. Lett.} \textbf{\bibinfo{volume}{112}},
  \bibinfo{pages}{080801} (\bibinfo{year}{2014}).

\bibitem[{\citenamefont{Herrera-Mart{\'\i}
  et~al.}(2015)\citenamefont{Herrera-Mart{\'\i}, Gefen, Aharonov, Katz, and
  Retzker}}]{herrera2015quantum}
\bibinfo{author}{\bibfnamefont{D.~A.} \bibnamefont{Herrera-Mart{\'\i}}},
  \bibinfo{author}{\bibfnamefont{T.}~\bibnamefont{Gefen}},
  \bibinfo{author}{\bibfnamefont{D.}~\bibnamefont{Aharonov}},
  \bibinfo{author}{\bibfnamefont{N.}~\bibnamefont{Katz}}, \bibnamefont{and}
  \bibinfo{author}{\bibfnamefont{A.}~\bibnamefont{Retzker}},
  \bibinfo{journal}{Phys. Rev. Lett.} \textbf{\bibinfo{volume}{115}},
  \bibinfo{pages}{200501} (\bibinfo{year}{2015}).

\bibitem[{\citenamefont{Gottesman}(2009)}]{gottesman2009introduction}
\bibinfo{author}{\bibfnamefont{D.}~\bibnamefont{Gottesman}},
  \bibinfo{journal}{Quantum Information Science and Its Contributions to
  Mathematics} \textbf{\bibinfo{volume}{68}}, \bibinfo{pages}{13}
  (\bibinfo{year}{2009}).

\bibitem[{\citenamefont{Gladchenko et~al.}(2009)\citenamefont{Gladchenko,
  Olaya, Dupont-Ferrier, Doucot, Ioffe, and
  Gershenson}}]{gladchenko2009superconducting}
\bibinfo{author}{\bibfnamefont{S.}~\bibnamefont{Gladchenko}},
  \bibinfo{author}{\bibfnamefont{D.}~\bibnamefont{Olaya}},
  \bibinfo{author}{\bibfnamefont{E.}~\bibnamefont{Dupont-Ferrier}},
  \bibinfo{author}{\bibfnamefont{B.}~\bibnamefont{Doucot}},
  \bibinfo{author}{\bibfnamefont{L.~B.} \bibnamefont{Ioffe}}, \bibnamefont{and}
  \bibinfo{author}{\bibfnamefont{M.~E.} \bibnamefont{Gershenson}},
  \bibinfo{journal}{Nature Physics} \textbf{\bibinfo{volume}{5}},
  \bibinfo{pages}{48} (\bibinfo{year}{2009}).

\bibitem[{\citenamefont{Reed et~al.}(2012)\citenamefont{Reed, DiCarlo, Nigg,
  Sun, Frunzio, Girvin, and Schoelkopf}}]{reed2012realization}
\bibinfo{author}{\bibfnamefont{M.}~\bibnamefont{Reed}},
  \bibinfo{author}{\bibfnamefont{L.}~\bibnamefont{DiCarlo}},
  \bibinfo{author}{\bibfnamefont{S.}~\bibnamefont{Nigg}},
  \bibinfo{author}{\bibfnamefont{L.}~\bibnamefont{Sun}},
  \bibinfo{author}{\bibfnamefont{L.}~\bibnamefont{Frunzio}},
  \bibinfo{author}{\bibfnamefont{S.}~\bibnamefont{Girvin}}, \bibnamefont{and}
  \bibinfo{author}{\bibfnamefont{R.}~\bibnamefont{Schoelkopf}},
  \bibinfo{journal}{Nature} \textbf{\bibinfo{volume}{482}},
  \bibinfo{pages}{382} (\bibinfo{year}{2012}).

\bibitem[{\citenamefont{C{\'o}rcoles et~al.}(2015)\citenamefont{C{\'o}rcoles,
  Magesan, Srinivasan, Cross, Steffen, Gambetta, and
  Chow}}]{corcoles2015demonstration}
\bibinfo{author}{\bibfnamefont{A.}~\bibnamefont{C{\'o}rcoles}},
  \bibinfo{author}{\bibfnamefont{E.}~\bibnamefont{Magesan}},
  \bibinfo{author}{\bibfnamefont{S.~J.} \bibnamefont{Srinivasan}},
  \bibinfo{author}{\bibfnamefont{A.~W.} \bibnamefont{Cross}},
  \bibinfo{author}{\bibfnamefont{M.}~\bibnamefont{Steffen}},
  \bibinfo{author}{\bibfnamefont{J.~M.} \bibnamefont{Gambetta}},
  \bibnamefont{and} \bibinfo{author}{\bibfnamefont{J.~M.} \bibnamefont{Chow}},
  \bibinfo{journal}{Nature communications} \textbf{\bibinfo{volume}{6}}
  (\bibinfo{year}{2015}).

\bibitem[{\citenamefont{Kelly et~al.}(2015)\citenamefont{Kelly, Barends,
  Fowler, Megrant, Jeffrey, White, Sank, Mutus, Campbell, Chen
  et~al.}}]{kelly2015state}
\bibinfo{author}{\bibfnamefont{J.}~\bibnamefont{Kelly}},
  \bibinfo{author}{\bibfnamefont{R.}~\bibnamefont{Barends}},
  \bibinfo{author}{\bibfnamefont{A.}~\bibnamefont{Fowler}},
  \bibinfo{author}{\bibfnamefont{A.}~\bibnamefont{Megrant}},
  \bibinfo{author}{\bibfnamefont{E.}~\bibnamefont{Jeffrey}},
  \bibinfo{author}{\bibfnamefont{T.}~\bibnamefont{White}},
  \bibinfo{author}{\bibfnamefont{D.}~\bibnamefont{Sank}},
  \bibinfo{author}{\bibfnamefont{J.}~\bibnamefont{Mutus}},
  \bibinfo{author}{\bibfnamefont{B.}~\bibnamefont{Campbell}},
  \bibinfo{author}{\bibfnamefont{Y.}~\bibnamefont{Chen}}, \bibnamefont{et~al.},
  \bibinfo{journal}{Nature} \textbf{\bibinfo{volume}{519}}, \bibinfo{pages}{66}
  (\bibinfo{year}{2015}).

\bibitem[{\citenamefont{Waldherr et~al.}(2014)\citenamefont{Waldherr, Wang,
  Zaiser, Jamali, Schulte-Herbr{\"u}ggen, Abe, Ohshima, Isoya, Du, Neumann
  et~al.}}]{waldherr2014quantum}
\bibinfo{author}{\bibfnamefont{G.}~\bibnamefont{Waldherr}},
  \bibinfo{author}{\bibfnamefont{Y.}~\bibnamefont{Wang}},
  \bibinfo{author}{\bibfnamefont{S.}~\bibnamefont{Zaiser}},
  \bibinfo{author}{\bibfnamefont{M.}~\bibnamefont{Jamali}},
  \bibinfo{author}{\bibfnamefont{T.}~\bibnamefont{Schulte-Herbr{\"u}ggen}},
  \bibinfo{author}{\bibfnamefont{H.}~\bibnamefont{Abe}},
  \bibinfo{author}{\bibfnamefont{T.}~\bibnamefont{Ohshima}},
  \bibinfo{author}{\bibfnamefont{J.}~\bibnamefont{Isoya}},
  \bibinfo{author}{\bibfnamefont{J.}~\bibnamefont{Du}},
  \bibinfo{author}{\bibfnamefont{P.}~\bibnamefont{Neumann}},
  \bibnamefont{et~al.}, \bibinfo{journal}{Nature}
  \textbf{\bibinfo{volume}{506}}, \bibinfo{pages}{204} (\bibinfo{year}{2014}).

\bibitem[{\citenamefont{Nigg et~al.}(2014)\citenamefont{Nigg, Mueller,
  Martinez, Schindler, Hennrich, Monz, Martin-Delgado, and
  Blatt}}]{nigg2014quantum}
\bibinfo{author}{\bibfnamefont{D.}~\bibnamefont{Nigg}},
  \bibinfo{author}{\bibfnamefont{M.}~\bibnamefont{Mueller}},
  \bibinfo{author}{\bibfnamefont{E.~A.} \bibnamefont{Martinez}},
  \bibinfo{author}{\bibfnamefont{P.}~\bibnamefont{Schindler}},
  \bibinfo{author}{\bibfnamefont{M.}~\bibnamefont{Hennrich}},
  \bibinfo{author}{\bibfnamefont{T.}~\bibnamefont{Monz}},
  \bibinfo{author}{\bibfnamefont{M.~A.} \bibnamefont{Martin-Delgado}},
  \bibnamefont{and} \bibinfo{author}{\bibfnamefont{R.}~\bibnamefont{Blatt}},
  \bibinfo{journal}{Science} \textbf{\bibinfo{volume}{345}},
  \bibinfo{pages}{302} (\bibinfo{year}{2014}).

\bibitem[{\citenamefont{Unden et~al.}(2016)\citenamefont{Unden,
  Balasubramanian, Louzon, Vinkler, Plenio, Markham, Twitchen, Stacey,
  Lovchinsky, Sushkov et~al.}}]{unden2016quantum}
\bibinfo{author}{\bibfnamefont{T.}~\bibnamefont{Unden}},
  \bibinfo{author}{\bibfnamefont{P.}~\bibnamefont{Balasubramanian}},
  \bibinfo{author}{\bibfnamefont{D.}~\bibnamefont{Louzon}},
  \bibinfo{author}{\bibfnamefont{Y.}~\bibnamefont{Vinkler}},
  \bibinfo{author}{\bibfnamefont{M.~B.} \bibnamefont{Plenio}},
  \bibinfo{author}{\bibfnamefont{M.}~\bibnamefont{Markham}},
  \bibinfo{author}{\bibfnamefont{D.}~\bibnamefont{Twitchen}},
  \bibinfo{author}{\bibfnamefont{A.}~\bibnamefont{Stacey}},
  \bibinfo{author}{\bibfnamefont{I.}~\bibnamefont{Lovchinsky}},
  \bibinfo{author}{\bibfnamefont{A.~O.} \bibnamefont{Sushkov}},
  \bibnamefont{et~al.}, \bibinfo{journal}{Phys. Rev. Lett.}
  \textbf{\bibinfo{volume}{116}}, \bibinfo{pages}{230502}
  (\bibinfo{year}{2016}).

\bibitem[{\citenamefont{Cohen et~al.}(2016)\citenamefont{Cohen, Pilnyak,
  Istrati, Retzker, and Eisenberg}}]{cohen2016demonstration}
\bibinfo{author}{\bibfnamefont{L.}~\bibnamefont{Cohen}},
  \bibinfo{author}{\bibfnamefont{Y.}~\bibnamefont{Pilnyak}},
  \bibinfo{author}{\bibfnamefont{D.}~\bibnamefont{Istrati}},
  \bibinfo{author}{\bibfnamefont{A.}~\bibnamefont{Retzker}}, \bibnamefont{and}
  \bibinfo{author}{\bibfnamefont{H.}~\bibnamefont{Eisenberg}},
  \bibinfo{journal}{Phys. Rev. A} \textbf{\bibinfo{volume}{94}},
  \bibinfo{pages}{012324} (\bibinfo{year}{2016}).

\bibitem[{\citenamefont{Slichter}(2013)}]{slichter2013principles}
\bibinfo{author}{\bibfnamefont{C.~P.} \bibnamefont{Slichter}},
  \emph{\bibinfo{title}{Principles of magnetic resonance}},
  vol.~\bibinfo{volume}{1} (\bibinfo{publisher}{Springer Science \& Business
  Media}, \bibinfo{year}{2013}).

\bibitem[{\citenamefont{Viola et~al.}(1999)\citenamefont{Viola, Knill, and
  Lloyd}}]{viola1999dynamical}
\bibinfo{author}{\bibfnamefont{L.}~\bibnamefont{Viola}},
  \bibinfo{author}{\bibfnamefont{E.}~\bibnamefont{Knill}}, \bibnamefont{and}
  \bibinfo{author}{\bibfnamefont{S.}~\bibnamefont{Lloyd}},
  \bibinfo{journal}{Physical Review Letters} \textbf{\bibinfo{volume}{82}},
  \bibinfo{pages}{2417} (\bibinfo{year}{1999}).

\bibitem[{\citenamefont{Bylander et~al.}(2011)\citenamefont{Bylander,
  Gustavsson, Yan, Yoshihara, Harrabi, Fitch, Cory, Nakamura, Tsai, and
  Oliver}}]{bylander2011noise}
\bibinfo{author}{\bibfnamefont{J.}~\bibnamefont{Bylander}},
  \bibinfo{author}{\bibfnamefont{S.}~\bibnamefont{Gustavsson}},
  \bibinfo{author}{\bibfnamefont{F.}~\bibnamefont{Yan}},
  \bibinfo{author}{\bibfnamefont{F.}~\bibnamefont{Yoshihara}},
  \bibinfo{author}{\bibfnamefont{K.}~\bibnamefont{Harrabi}},
  \bibinfo{author}{\bibfnamefont{G.}~\bibnamefont{Fitch}},
  \bibinfo{author}{\bibfnamefont{D.~G.} \bibnamefont{Cory}},
  \bibinfo{author}{\bibfnamefont{Y.}~\bibnamefont{Nakamura}},
  \bibinfo{author}{\bibfnamefont{J.~S.} \bibnamefont{Tsai}}, \bibnamefont{and}
  \bibinfo{author}{\bibfnamefont{W.~D.} \bibnamefont{Oliver}},
  \bibinfo{journal}{Nature Physics} \textbf{\bibinfo{volume}{7}},
  \bibinfo{pages}{565} (\bibinfo{year}{2011}).

\bibitem[{\citenamefont{Yan et~al.}(2015)\citenamefont{Yan, Gustavsson, Kamal,
  Birenbaum, Sears, Hover, Gudmundsen, Yoder, Orlando, Clarke
  et~al.}}]{yan2015flux}
\bibinfo{author}{\bibfnamefont{F.}~\bibnamefont{Yan}},
  \bibinfo{author}{\bibfnamefont{S.}~\bibnamefont{Gustavsson}},
  \bibinfo{author}{\bibfnamefont{A.}~\bibnamefont{Kamal}},
  \bibinfo{author}{\bibfnamefont{J.}~\bibnamefont{Birenbaum}},
  \bibinfo{author}{\bibfnamefont{A.}~\bibnamefont{Sears}},
  \bibinfo{author}{\bibfnamefont{D.}~\bibnamefont{Hover}},
  \bibinfo{author}{\bibfnamefont{T.}~\bibnamefont{Gudmundsen}},
  \bibinfo{author}{\bibfnamefont{J.}~\bibnamefont{Yoder}},
  \bibinfo{author}{\bibfnamefont{T.}~\bibnamefont{Orlando}},
  \bibinfo{author}{\bibfnamefont{J.}~\bibnamefont{Clarke}},
  \bibnamefont{et~al.}, \bibinfo{journal}{arXiv preprint arXiv:1508.06299}
  (\bibinfo{year}{2015}).

\bibitem[{\citenamefont{Lindblad}(1976)}]{lindblad76}
\bibinfo{author}{\bibfnamefont{G.}~\bibnamefont{Lindblad}},
  \bibinfo{journal}{Communications in Mathematical Physics}
  \textbf{\bibinfo{volume}{48}}, \bibinfo{pages}{119} (\bibinfo{year}{1976}).

\bibitem[{\citenamefont{Hornberger}(2009)}]{hornberger2009introduction}
\bibinfo{author}{\bibfnamefont{K.}~\bibnamefont{Hornberger}}, in
  \emph{\bibinfo{booktitle}{Entanglement and Decoherence}}
  (\bibinfo{publisher}{Springer}, \bibinfo{year}{2009}), pp.
  \bibinfo{pages}{221--276}.

\bibitem[{\citenamefont{Gardiner and Zoller}(2004)}]{GZ01b}
\bibinfo{author}{\bibfnamefont{C.~W.} \bibnamefont{Gardiner}} \bibnamefont{and}
  \bibinfo{author}{\bibfnamefont{P.}~\bibnamefont{Zoller}},
  \emph{\bibinfo{title}{Quantum Noise}} (\bibinfo{publisher}{Springer, Berlin},
  \bibinfo{year}{2004}).

\bibitem[{\citenamefont{Hein et~al.}(2005)\citenamefont{Hein, D{\"u}r, and
  Briegel}}]{hein2005entanglement}
\bibinfo{author}{\bibfnamefont{M.}~\bibnamefont{Hein}},
  \bibinfo{author}{\bibfnamefont{W.}~\bibnamefont{D{\"u}r}}, \bibnamefont{and}
  \bibinfo{author}{\bibfnamefont{H.-J.} \bibnamefont{Briegel}},
  \bibinfo{journal}{Physical Review A} \textbf{\bibinfo{volume}{71}},
  \bibinfo{pages}{032350} (\bibinfo{year}{2005}).

\end{thebibliography}

\end{document}